\documentstyle[12pt]{article}

 at 10truept

\def\be{\beta}
\def\ga{\gamma}
\def\de{\delta}
\def\ep{\epsilon}

\def\ka{\kappa}
\def\la{\lambda}

\def\si{\sigma}

\def\ph{\phi}

\def\ps{\psi}

\def\De{\Delta}

\def\cl{{\cal L}}

\def\half{{\textstyle{1\over 2}}}
\def\frac#1#2{{\textstyle{{#1}\over {#2}}}}

\def\lsim{\mathrel{\rlap{\lower4pt\hbox{\hskip1pt$\sim$}}
    \raise1pt\hbox{$<$}}}
\def\gsim{\mathrel{\rlap{\lower4pt\hbox{\hskip1pt$\sim$}}
    \raise1pt\hbox{$>$}}}
\def\sqr#1#2{{\vcenter{\vbox{\hrule height.#2pt
         \hbox{\vrule width.#2pt height#1pt \kern#1pt
         \vrule width.#2pt}
         \hrule height.#2pt}}}}

\def\Im{\hbox{Im}\,}

\def\lrprtmu{\stackrel{\leftrightarrow}{\partial_\mu}}
\def\lrprtnu{\stackrel{\leftrightarrow}{\partial^\nu}}

\newcommand{\beq}{\begin{equation}}
\newcommand{\eeq}{\end{equation}}
\newcommand{\bea}{\begin{eqnarray}}
\newcommand{\eea}{\end{eqnarray}}
\newcommand{\rf}[1]{(\ref{#1})}

\renewenvironment{thebibliography}[1]
 { \rm
   \begin{list}{\arabic{enumi}.}
    {\usecounter{enumi} \setlength{\parsep}{0pt}
     \setlength{\itemsep}{3pt} \settowidth{\labelwidth}{#1.}
     \sloppy
    }}{\end{list}}

\begin{document}

\begin{flushright}
{IUHET 386\\}
{March 1998\\}
\end{flushright}
\vglue 1.0 truein 

\begin{flushleft}
{\bf TESTING A CPT- AND LORENTZ-VIOLATING EXTENSION\\
  OF THE STANDARD MODEL%
\footnote{
Presented at Orbis Scientiae 1997-II, 
Miami, Florida, December 1997}
\\}
\end{flushleft}

\vglue 0.8cm
\begin{flushleft}
{\hskip 1 truein
V. Alan Kosteleck\'y
\\}
\bigskip
{\hskip 1 truein
Physics Department\\}
{\hskip 1 truein
Indiana University\\}
{\hskip 1 truein
Bloomington, IN 47405\\}
{\hskip 1 truein
U.S.A.\\}
\end{flushleft}

\vglue 0.8cm

\noindent
{\bf INTRODUCTION AND BACKGROUND}
\vglue 0.4 cm 

The standard model of particle physics is invariant under
a variety of continuous symmetry operations,
including translations, 
Lorentz transformations,
and gauge transformations.
The model is also invariant under the action 
of the product CPT of charge conjugation C, 
parity reflection P, and time reversal T.
Indeed,
CPT symmetry is known to be a characteristic
of all local relativistic field theories of point particles
\cite{sachs}.
It has been experimentally tested to high accuracy 
in a variety of situations
\cite{pdg}.
The general validity of CPT symmetry for particle theories
and the existence of high-precision tests
means CPT breaking is an interesting candidate experimental signal 
for new physics beyond the standard model,
such as might emerge in the context of string theory
\cite{kp1,kp2,kp3}.

In a talk 
\cite{os1}
delivered at the previous meeting in this series
(Orbis Scientiae 1997-I),
I discussed the possibility 
that CPT and Lorentz symmetry
might be broken in nature by effects
emerging from a fundamental theory beyond the standard model.
String theory,
which currently represents the most promising framework 
for a consistent quantum theory of gravity 
incorporating the known particles and interactions,
is a candidate theory in which effects of this type might occur. 
The point is that strings are extended objects,
so the standard axioms underlying proofs of CPT invariance 
are inappropriate.
In fact,
it is known that spontaneous CPT and Lorentz violation 
can occur in the context of string theory
\cite{kp1,ks}.

If the fundamental theory has Lorentz and CPT symmetry
and is naturally formulated in more than four spacetime dimensions,
then some kind of spontaneous breaking 
of the higher-dimensional Lorentz group 
presumably must occur to produce an effective low-energy theory
with only four macroscopic dimensions.
This situation exists for some string theories, 
for example.
An interesting issue is whether the spontaneous breaking
generates apparent Lorentz and CPT violation
in our four spacetime dimensions.
It might seem natural for this to happen,
since there is no evident reason why
four dimensions would be preferred 
in the higher-dimensional theory.
However, 
no experimental evidence exists for Lorentz or CPT breaking,
so if it occurs it must be highly suppressed
at the level of the standard model.
If the standard model is regarded as an effective low-energy theory
emerging from a realistic string theory,
then the natual dimensionless suppression factor 
for observable Lorentz or CPT violation
would be the ratio $r$ of the low-energy scale to the Planck scale,
$r \sim 10^{-17}$.
Relatively few experiments would be sensitive to such effects.

In the previous talk
\cite{os1},
I outlined the low-energy description of effects 
from spontaneous Lorentz and CPT breaking 
in an underlying theory.
At this level,
the potentially observable Lorentz and CPT violations
appear merely as consequences of the vacuum structure,
so many desirable properties of Lorentz-invariant models
are maintained.
The low-energy theory acquires additional terms with a generic form
\cite{kp2,kp3}.
More specifically,
at the level of the standard model,
refs.\ \cite{cksm1,cksm2}
have identified the most general terms 
that can arise from spontantaneous Lorentz violation
(both with and without CPT breaking) 
while maintaining SU(3) $\times$ SU(2) $\times$ U(1) 
gauge invariance and power-counting renormalizability.
The existence of this explicit extension of the standard model
offers the possibility of quantitative investigations
of a variety of experimental signals 
for apparent Lorentz and CPT breaking.

Some possible consequences of the additional terms 
in the standard-model extension
were presented at the previous meeting 
\cite{os1}.
Among the most interesting quantitative tests of CPT
are experiments with neutral-meson oscillations
in the $K$ system
\cite{kp1,kp2,kp3,k}, 
the two $B$ systems
\cite{kp3,ck1,kv},
and the $D$ system
\cite{kp3,ck2}. 
Implications of CPT violation in other contexts,
such as baryogenesis \cite{bckp},
were also described.

In the present talk,
I provide an update of some developments
that have occurred in the months since
the previous meeting.
Possible experimental tests of
the QED limit of the standard-model extension
have been examined 
\cite{cksm2,bkr1,osbkr}.
The sensitivity of tests of CPT violation 
in neutral-meson systems has been investigated
\cite{k},
and the first experimental results have been obtained  
constraining CPT violation in the neutral-$B$ system
\cite{opal,delphi}.

\vglue 0.6 cm 
\noindent
{\bf EXTENDED QUANTUM ELECTRODYNAMICS}
\vglue 0.4 cm 

The general Lorentz-violating extension of the standard model 
(including terms with and without CPT violation),
explicitly given in refs.\ \cite{cksm1,cksm2},
follows from imposing two requirements.
One is that the form of the additional terms
must be compatible with an origin 
from spontaneous Lorentz breaking in an underlying theory.
The other is that the usual properties of
SU(3) $\times$ SU(2) $\times$ U(1) gauge invariance 
and power-counting renormalizability 
must be maintained.
These criteria suffice to keep relatively small
the number of new terms in the action.
A framework for treating the implications
of apparent Lorentz and CPT violation has also 
been presented in the above works.

One limit of this extended standard model
is an extension of quantum electrodynamics (QED)
\cite{cksm2}.
This is of particular interest
because QED is a well-established theory
for which numerous experimental tests exist.
Here,
I give only the lagrangian for the extended 
theory of photons, electrons, and positrons,
which has a relatively simple form.

The usual lagrangian is 
\beq
\cl^{\rm QED} =
\overline{\ps} \ga^\mu (\half i \lrprtmu e - q A_\mu ) \ps 
- m \overline{\ps} \ps 
- \frac 1 4 F_{\mu\nu}F^{\mu\nu}
\quad .
\label{a}
\eeq
The CPT-violating terms are
\beq
\cl^{\rm CPT}_{e} =
- a_{\mu} \overline{\ps} \ga^{\mu} \ps 
- b_{\mu} \overline{\ps} \ga_5 \ga^{\mu} \ps \quad ,
$$
$$
\cl^{\rm CPT}_{\ga} =
\half (k_{AF})^\ka \ep_{\ka\la\mu\nu} A^\la F^{\mu\nu}
\quad .
\label{b}
\eeq
The Lorentz-violating but CPT-preserving terms are
\beq
\cl^{\rm Lorentz}_{e} = 
c_{\mu\nu} \overline{\ps} \ga^{\mu} 
(\half i \lrprtnu - q A^\nu ) \ps 
+ d_{\mu\nu} \overline{\ps} \ga_5 \ga^\mu 
(\half i \lrprtnu - q A^\nu ) \ps 
- \half H_{\mu\nu} \overline{\ps} \si^{\mu\nu} \ps 
$$
$$
\cl^{\rm Lorentz}_{\ga} =
-\frac 1 4 (k_F)_{\ka\la\mu\nu} F^{\ka\la}F^{\mu\nu}
\quad .
\label{c}
\eeq
The coefficients of the various terms can be regarded
as Lorentz- and CPT-violating couplings.
The reader is directed to refs.\ \cite{cksm1,cksm2}
for details of notations and conventions
as well as for more information about the various terms, 
including issues such as the effect of field redefinitions
and the possibility of other couplings.

As mentioned above,
many conventional tests of Lorentz and CPT symmetry
are expected to be insensitive to effects 
from the additional terms in the extension of QED
because of the expected small size of the couplings.
Nonetheless,
certain kinds of experiment can provide constraints.

First, 
consider the fermion sector.
One important class of tests 
consists of Penning-trap experiments
measuring anomaly and cyclotron frequencies
with exceptional precision
\cite{ge,qme,gp,qmp}.
These have been investigated in the present context in 
refs.\ \cite{bkr1},
where possible signals are identified,
appropriate figures of merit are introduced,
and estimates are given of limits on Lorentz and CPT violation
that would be attainable in present and future experiments.
A summary of the results of these works
can be found in a separate contribution
to the present volume
\cite{osbkr}. 
As one example,
the spacelike components of the coefficient $b_\mu$
can be bounded by experiments
comparing the anomalous magnetic moments
of the electron and positron.
The associated figure of merit for CPT violation
could be constrained to about one part in $10^{20}$.
This is comparable to the ratio 
of the electron mass to the Planck scale
at which suppressed but observable effects 
from an underlying theory might be expected.
Some interesting constraints on a subset of couplings 
in the fermion sector of extended QED might also arise 
from high-precision experiments of various other kinds,
including clock-comparison tests
\cite{cc}.

Next,
consider the photon sector of the QED extension
\cite{cksm1,cksm2}.
The CPT-breaking term
with coefficient $(k_{AF})_\mu$
has theoretical difficulties
in that the associated canonical energy can be 
negative and arbitrarily large.
This suggests that the coefficient should vanish,
which in turn provides an interesting 
theoretical consistency check of the model.
The point is that,
even if this coefficient vanishes at tree level,
it would typically be expected to acquire radiative corrections 
involving CPT-breaking couplings from the fermion sector,
which in the present context could cause difficulty
with the positivity of the theory.
However,
it has been shown 
\cite{cksm2}
that no such radiative corrections arise
in the context of the standard-model extension described above.
At the experimental level,
limits from cosmological birefringence
restrict the components of 
$(k_{AF})_\mu$ to $\lsim 10^{-42}$ GeV
\cite{cfj},
although there exist disputed claims 
\cite{nr}
for a nonzero effect corresponding 
to $|\vec k_{AF}|\sim 10^{-41}$ GeV.

In contrast,
a nonzero contribution from 
the CPT-preserving, Lorentz-breaking term 
in the photon sector of the QED extension
would maintain the positivity 
of the total canonical energy density
and appears to be theoretically allowed
\cite{cksm2}.
Moreover,
even if the coefficients 
$(k_F)_{\ka\la\mu\nu}$ 
vanish at tree level,
one-loop corrections from the fermion sector are induced.
It is therefore of interest to examine 
possible experimental constraints on this type of term.
One irreducible component of 
$(k_F)_{\ka\la\mu\nu}$ 
is rotation invariant
and can be bounded to $\lsim 10^{-23}$ 
by the existence of cosmic rays 
\cite{g}
or by other tests.
The remaining components violate rotation invariance 
and might in principle be bounded by cosmological birefringence.
The attainable bounds are substantially weaker than those
discussed above for the CPT-breaking term because, 
unlike $(k_{AF})_\mu$,
the coefficients $(k_F)_{\ka\la\mu\nu}$
are dimensionless and so are suppressed 
by the energy scale of the radiation involved.
Further details about the photon sector of the QED extension 
can be found in ref.\ \cite{cksm2}.

\vglue 0.6 cm 
\noindent
{\bf NEUTRAL-MESON OSCILLATIONS}
\vglue 0.4 cm 

Since the last meeting in this series,
there have been several developments
concerning the possibility of testing 
the standard-model extension 
using neutral-meson oscillations.
In what follows,
a generic neutral meson is denoted by $P$,
where $P \equiv K$, $D$, $B_d$, or $B_s$.

Interferometry with $P$ mesons
can involve two types of (indirect) CP violation:
T violation with CPT invariance,
or CPT violation with T invariance.
These are phenomenologically described by
complex parameters $\ep_P$ and $\de_P$,
respectively,
that are introduced in the effective hamiltonian 
for the time evolution of a neutral-meson state.
Within the context of the standard-model extension,
it can be shown that the CPT-violating parameter $\de_P$ 
depends only on one of the types of additional coupling 
\cite{k}.
Only CPT-violating terms in the lagrangian 
of the form
$- a^q_{\mu} \overline{q} \ga^\mu q$
are relevant,
where $q$ is a quark field
and the coupling $a^q_{\mu}$ is constant in spacetime 
but depends on the quark flavor $q$.
It is also noteworthy that 
the parameters $\de_P$ are the only quantities 
known to be sensitive to the couplings $a_\mu$.

To define $\de_P$,
one must work in a frame comoving with the $P$ meson.
It can be shown that the CPT and Lorentz breaking
introduces a dependence of $\de_P$ 
on the boost and orientation of the meson.
Let the $P$-meson four-velocity be
$\be^\mu \equiv \ga(1,\vec\be)$.
Then,
at leading order in all Lorentz-breaking couplings 
in the standard-model extension,
$\de_P$ is given by
\cite{k}
\beq
\de_P \approx i \sin\hat\ph \exp(i\hat\ph) 
\ga(\De a_0 - \vec \be \cdot \De \vec a) /\De m
\quad .
\label{e}
\eeq
In this expression,
$\De a_\mu \equiv a_\mu^{q_2} - a_\mu^{q_1}$,
where $q_1$ and $q_2$ denote the valence-quark flavors 
in the $P$ meson.
The quantity $\hat\ph$ is given by
$\hat\ph\equiv \tan^{-1}(2\De m/\De\ga)$,
where $\De m$ and $\De \ga$
are the mass and decay-rate differences
between the $P$-meson eigenstates,
respectively.
Note that a subscript $P$ is suppressed on all 
variables on the right-hand side of Eq.\ \rf{e}.

One implication of the above results for experiment is 
a proportionality between the real and imaginary 
components of $\de_P$
\cite{kp2,kp3}.
A second is the possibility of a variation of the magnitude
of $\de_P$ with $P$,
arising from the 
flavor dependence of the couplings $a_\mu^q$
\cite{kp3}.
Other implications arise 
from the momentum and orientation dependences in Eq.\ \rf{e},
which offer the possibility of striking signals
for Lorentz and CPT breaking
\cite{k}.
The momentum and orientation dependences 
also imply an enhanced signal for boosted mesons
and suggest that published bounds on $\de_P$ 
from distinct experiments could represent 
different CPT sensitivities.
Experiments involving highly boosted mesons,
such as the $K$-system experiment E773 at Fermilab
\cite{e773},
would be particularly sensitive to Planck-scale effects.

The tightest neutral-meson bounds on CPT violation
at present are from experiments with the neutral-$K$ system.
The possibility exists that relatively large CPT violation
might occur in the behavior of heavier neutral mesons.
At the time of the previous meeting in this series,
no bounds existed on CPT violation in the $D$ or $B$ systems.
My talk at that meeting
\cite{os1}
emphasized that sufficient data already existed 
to place bounds on CPT violation in the $B_d$ system
\cite{kv}.
Since then,
two experimental groups at CERN
have performed the suggested measurement.
The OPAL collaboration  
has published the result
\cite{opal}
$\Im\de_{B_d} = -0.020 \pm 0.016 \pm 0.006$,
while the DELPHI collaboration
has released a preliminary measurement 
\cite{delphi}
$\Im\de_{B_d} = -0.011 \pm 0.017 \pm 0.005$.
Other analyses of CPT violation in heavy-meson systems
are presently underway.

\vglue 0.6 cm
\noindent
{\bf ACKNOWLEDGMENTS}
\vglue 0.4 cm

I thank Orfeu Bertolami, Robert Bluhm, Don Colladay, 
Rob Potting, Neil Russell, Stuart Samuel, 
and Rick Van Kooten for collaborations.
This work is supported in part
by the United States Department of Energy 
under grant number DE-FG02-91ER40661.

\vglue 0.6 cm
\noindent
{\bf REFERENCES}
\vglue 0.4 cm

\end{document}